\documentclass[11pt,english]{article} 
\usepackage{times,graphicx,graphics,pstricks,pst-node,pst-coil,pst-grad}
\newpsobject{showgrid}{psgrid}{subgriddiv=1,griddots=10,gridlabels=6pt}
\def\be{\begin{equation}} 
\def\eq{\end{equation}} 

\usepackage{graphicx} 
\usepackage{amsmath}
\usepackage{amssymb}
\usepackage{enumerate}

\def\[{\begin{equation}}
\def\]{\end{equation}}

\begin{document}  
\title{Simple circuit theory and the solution of two electricity problems from the Victorian Age}  
\author{A C Tort \footnote{e-mail: tort@if.ufrj.br.}\\
Departamento de F\'{\i}sica Te\'{o}rica - Instituto de F\'{\i}sica \\
Universidade Federal do Rio de Janeiro\\ Caixa Postal 68.528; CEP 21941-972 Rio
de Janeiro, Brazil} 
\maketitle
\begin{abstract} 
Two problems from the Victorian Age,  the subdivision of light and the determination of the leakage point in an undersea telegraphic cable are discussed and suggested as a concrete illustrations of the relationships  between textbook physics and the real world.  Ohm's law and simple algebra are the only tools we need to discuss them in the classroom.
\end{abstract}
\newpage
\section{Introduction}  
Some time ago, the present author had the opportunity of reading Paul J. Nahin's \cite{PN} fascinating biography of the Victorian physicist and electrician Oliver Heaviside (1850-1925).  Heaviside's scientific life unrolls against a background of theoretical and technical challenges that the scientific and technological developments fostered by the Industrial Revolution presented to engineers and physicists of those times. It is a time where electromagnetic theory as formulated by James Clerk Maxwell (1831-1879) was understood by only a small group of men, Lodge, FitzGerald and Heaviside, among others, that had the mathematical sophistication and imagination to grasp the meaning and take part in the great Maxwellian synthesis. Almost all of the electrical engineers, or electricians as they were called at the time, considered themselves as \lq\lq practical men\rq\rq, \hskip 0.1cm which effectively meant that most of them had a working knowledge of the electromagnetic phenomena spiced up with bits of electrical theory, to wit,  Ohm's law and the Joule effect. Two problems belonging to that time are described by Nahin \cite{PN} and briefly discussed as end of chapter technical notes. Those two problems called the present author's attention for their potentiality as simple pedagogical examples capable of establishing strong links between textbook physics and the real world of economical and industrial affairs. 

The first problem concerns the possibility of replacing domestic lightning, that was then provided by the burning of gases, by d.c. electricity combined with the novel (at the time) incandescent light bulb patented by Thomas Alva Edison (1847-1931) in 1880. This is the problem of the subdivision of light or current. The second problem concerns the ingenious solution found by Oliver Heaviside to the problem of finding the location of the leakage point of an undersea telegraphic cable, at the time the ultimate means to convey information from one point to another. As the engineers of the Victorian Age, we will make use of Ohm's law, the Joule effect, and simple algebra to discuss those two problems.

\noindent
As mentioned before, those two examples are a byproduct of the author's reading of Nahin's \cite{PN} biography of Oliver Heaviside, and the main purpose of this paper is to call the attention of physics teachers and students to their pedagogical possibilities as classroom examples or homework assignment, or just collateral reading. 
 
\section{The subdivision of light}

In the years 1870, the only sources of light apart from candles and oil lamps, were the burning of gases and the electric arc. Oil lamps were fueled with olive oil, fish oil, whale oil, and sesame oil. Candles were made of beewax and fat. Gas lightning made use of natural gas or charcoal gas. With the practical applications of electricity, lightning with arc lamps became a reality in many european cities. Arc lamps, together with gas lamps, were employed in street illumination, but were not appropriate for domestic lightning\footnote{The history of public illumination is a fascinating subject. The administrators of the main european cities such as London, Paris and Madrid, soon noticed a strong correlation between the illumination of streets and public spaces and the diminution in the crime rates.  London in 1812 and Paris in 1820 adopted gas illumination. }.  For domestic illumination, gas lighting was more economical and convenient. The invention of the incandescent light bulb changes it all. Edison's  light bulb had a great economical potential and the stock market reacted swiftly. The price of the gas company stocks suffered a marked devaluation.
%
\noindent
The Welsh engineer William Henry Preece (1834-1913) tried to calm down the stockholders by producing a technical analysis whose sole purpose was to demonstrate the practical impossibility of having several incandescent light bulbs functioning at the same time with the sufficient efficaciousness necessary to illuminate an entire domestic residence. Preece, one of the most important electrical engineers of the Victorian Age, was the perfect example  of the so called \lq\lq practical man.\rq\rq\hskip 0.1cm  He had a profound dislike of the mathematisation of the electromagnetism, a dislike that he often manifested explicitly. On the problem of the subdivision of light (or current), as the problem was known at the time, he stated, as quoted in  \cite{PN}:  \lq\lq the extensive subdivision of the light must be ranked with perpetual motion, squaring of the circle, and the transmutation of metals.\rq \rq  \hskip 0.10cm and concluded: \lq\lq electricity cannot supplant gas for domestic purposes.\rq\rq\hskip 0.10cm  Preece also never understood the importance of Maxwell's theory of electromagnetic phenomena.  In other aspects, however, Preece was a competent engineer and administrator.

\begin{figure}[!h]

\begin{center}

\begin{pspicture}(-6,-2)(6,4)


\psset{arrowsize=0.2 2}

\pszigzag[coilarm=1.0, coilheight=0.15, linewidth=0.35mm]{-}(-4, 2.5)(0, 2.5)

\pszigzag[coilarm=1.0, coilheight=0.15, linewidth=0.35mm]{-}(0, 2.5)(4, 2.5)

\pszigzag[coilarm=1.0, coilheight=0.15, linewidth=0.35mm]{-}(0,0)(4, 0)


\rput(-2, 3.35){$r$}

\rput(2, 3.35){$r^{\,\prime}$}

\rput(2,-1){$R_{\mbox{\tiny eq}}$}

\psline[linewidth=0.35mm](-4,2.5)(-4,0)

\psline[linewidth=0.35mm](4,2.5)(4,0)

\psline[linewidth=0.85mm](-2,1)(-2,-1)

\psline[linewidth=0.85mm](-1.75,0.5)(-1.75,-0.5)

\psline[linewidth=0.35mm](-4,0)(-2,0)

\psline[linewidth=0.35mm](0,0)(-1.75,0)

\rput(-2, -1.35){$\mathcal{E}$}

\end{pspicture}

\caption{ The equivalent resistance represent the array of light bulbs. }
\label{circuit1}
\end{center}

\end{figure}
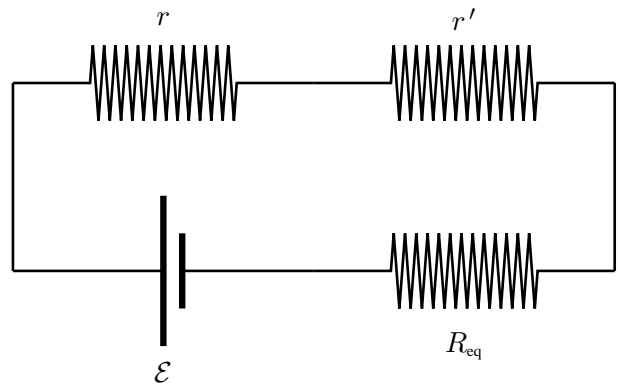

With pedagogical purposes in mind let us retrace with a certain amount of  detail Preece's steps.  The starting point will be a simple d.c. circuit\footnote{The transmission of electric power by means of an a.c. circuit, invented by Nicola Tesla and patronised by George Westinghouse was more efficient, but Edison had patented a system of distribution of electric power based on d.c. circuits.} made with a voltage source, two resistors connected in series, one modelling  the internal resistance of the source and the other one the resistance of the wiring. To these circuit elements we will add a finite number of identical resistors that can be interconnected in series or parallel. These resistors will model the light bulbs that  provide the domestic lightning. Let us denote by 
 $r$, the internal resistance, and by $r^{\,\prime}$, the resistance of the wiring. The resistance of a light bulb will be denoted by $R$ and the voltage of the source by  $\mathcal{E}$. The array of light bulbs, in series or in parallel, is represented by an equivalent $R_{\mbox{\tiny eq}}$,  see Fig.\ref{circuit1}. 
%
\noindent  Let us consider in first place the case in which the light bulbs are interconnected in series. Let $n$, a positive integer,  denote the number of light bulbs.  In this case the array can be represented by $R_{\mbox{\tiny eq}}=nR$. According to Ohm's law, the current through the circuit is
 
\[\label{corrente} i=\frac{\mathcal{E}}{\left(r+r^{\,\prime}+nR\right)}\,, \]
\noindent
consequently, due to the Joule effect, the power dissipated as light and heat by this array of light bulbs is

\[ P_{\mbox{\tiny series}}=i^2\,R_{\mbox{\tiny eq}}=\frac{\mathcal{E}^2\,nR}{\left(r+r^{\,\prime}+nR\right)^2}\, .\]
If, on the other hand, the light bulbs are interconnected in parallel, the equivalent resistance is $R_{\mbox{\tiny eq}}=R/n$.  The new current is 

\[\label{corrente2} i^{\,\prime}=\frac{\mathcal{E}}{\left(r+r^{\,\prime}+R/n\right)}\,, \]
and the power dissipated by the parallel array is 

\[P_{\mbox{\tiny parallel}}=\frac{\mathcal{E}^2\,R/n}{\left(r+r^{\,\prime}+R/n\right)^2} .\]

\noindent The power dissipated by a single light bulb in the first case is

\[P^{\mbox{\,\tiny bulb}}_{\mbox{\tiny series}}=\frac{\mathcal{E}^2\,R}{\left(r+r^{\,\prime}+nR\right)^2} , \]
 and in the second case
 
 \[\label{case2} P^{\mbox{\,\tiny bulb}}_{\mbox{\tiny parallel}}=\frac{\mathcal{E}^2\,R/n^2}{\left(r+r^{\,\prime}+R/n\right)^2} .\]

The controversial part of Preece's analysis begins here. Preece now supposes that when the number of light bulbs is very large,  ($n \gg 1$),  the conditions $r+r^{\,\prime}\ll nR$,  for the first case, the one in which the light bulbs are connected in series, and
$r+r^{\,\prime}\gg R/n$, for the case in which the light bulbs are connected in parallel, holds.  It follows that in this limit, in the first case 

\[  P^{\mbox{\,\tiny bulb}}_{\mbox{\tiny series}}\approx \frac{\mathcal{E}^ 2}{n^2\,R} , \]
and in the second case

\[ P^{\mbox{\,\tiny bulb}}_{\mbox{\tiny parallel}}\approx \frac{\mathcal{E}^ 2\,R}{n^2\,\left(r+r^{\,\prime}\right)^2 } , \]
Preece concludes that for both cases the power dissipated by a single light bulb varies with the reciprocal of the square of the number of bulbs, hence, for both cases, the illumination that a single bulb yields is negligible. Therefore, it is not reasonable to change the domestic lightning system from gas lamps to incandescent light bulbs powered by electricity. Notice that in the second case, the parallel one, Preece supposes that the combination $r+r^{\,\prime}$ \ is of the same order of magnitude as $R$. 

Preece's analysis is only partially correct. In Edison's original idea, the combination $r+r^{\,\prime}$ is a fraction of a ohm and the resistance of the light bulb approximately equal to $200\,\Omega$.  If, for instance, we assume  $r+r^{\,\prime}\approx 0,1\,\Omega$, is easy to see that in order to have $r+r^{\,\prime}$ of the same order of magnitude as $R$ we will need  $2000$ light bulbs! Evidently this is not the case for an ordinary residence. This also means that the correct assumption to consider in the parallel case is $r+r^{\,\prime}\ll R/n$, for a reasonable $n$. In this case equation (\ref{case2}) leads to the result

\[P^{\mbox{\tiny bulb}}_{\mbox{\tiny parallel}}\approx \frac{\mathcal{E}^ 2}{R} ,\]
that is, all light bulbs shine with the same intensity. In the other case, \textit{i.e.}, the case in which the light bulbs are interconnected in series, Preece's analysis is correct. 

Later, Preece admitted publicly his error and played a role in the implementation of electric lightning of the City of London \cite{Bourne}. In January 1882, Edison inaugurated the first electric power station of London, at the Holborn Viaduct. The system worked with direct current and provided electric power for the illumination of the streets and domestic residences near the station. Proximity to the power station was one of the weak points of Edison's system. Though fiercely attacked by Edison in an episode known as \lq\lq the war of currents,\rq\rq \hskip 0.1cm the d.c. based system would be soon replaced by the more efficient a.c. based one.
%
%
\section{ Locating the leakage point }
One of the important achievements  of the 19th century is the undersea telegraphic cable which made possible to transmit messages using a code that could be translated into electric impulses. Among several problems related to the operation of the telegraphic system there was the problem of the determination of the point of current leakage, a difficult task  because the cable lay on the bottom of the ocean. Oliver Heaviside (1850-1925) applied a method developed by the French telegraph engineer  \'Edouard Ernest Blavier (1826-1887) in order to locate the leakage point in the undersea cable  joining Newbiggin-by-the-Sea in England to Sondervig in Denmark. 


Heaviside's behaviour some times strange, antisocial and even bizarre did not prevented him of becoming one of the great names in the history of electromagnetic theory. Heaviside's contributions to this field are many and varied, for instance, the operational calculus and the theoretical  basis of cable telegraphy. Except for the taking into account of the luminiferous ether concept, Heaviside and the outstanding German physicist  Heinrich Rudolph Hertz (1857-1894) are responsible for the modern form in which we teach Maxwell's equations to our students \cite{Darrigol}. Though this is well known by historians of science, few textbooks mention it. But there are exceptions, see, for example,  \cite{Lorrain}.

\begin{figure}[!h]
\begin{center}
\begin{pspicture}(-3,-3)(3,2)
\dotnode(-3.0,0.0){A}
\dotnode(3.0,0.0){B} \ncline[linewidth=0.3mm]{A}{B}
\psline[linewidth=0.2mm]{->}(0,0)(3,-2) \rput(-1.5,0.25){$x$}
\rput(1.5,0.25){$a-x$} \rput(1.25,-1.5){$y$}
\psdots(0,0)
\end{pspicture}
\caption{Heaviside's sketch of the problem.}
\label{sketch}
\label{cabosub}
\end{center}
\end{figure}
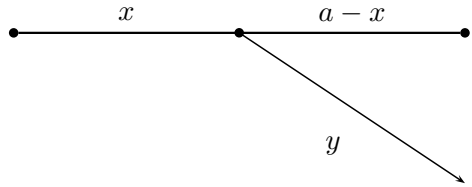

One of Heavisde's notebooks shows that he applied himself to the solution of the problem of locating the leakage point on 16 January 1871 \cite{PN}. Again with pedagogical purposes in mind, let us take a closer look at Heaviside's, then a 21years old youngman, approach to the problem.  As was usual at the time, Heaviside's analysis presupposes that it is possible to apply the most basic elements  of the d.c. circuit theory. In particular, by treating the cable as an ordinary ohmic conductor through which a constant current flows, Heaviside realises that he can make use of the formula relating the resistance $R$ of a segment of the conductor to its length $\ell$ \cite{Tipler}  

\[
R = \rho \,\frac{\ell }{A},
\]
where $\rho$ is the resistivity of the conductor and $A$, the cross section area.  Defining the resistivity per unit area, or resistance per unit length,  by 

\[r=\frac{\rho}{A},\]
\noindent we have

\[R=r\,\ell .\]
We see that the concept of resistance is intertwined with that of length and the measuring of the resistance is virtually the same as the measuring of length. Heaviside denotes the cable resistance from end to end \textit{without current losses} by $x$; the cable resistance between Newbiggin-by-the Sea and the leakage point by $x$, and the resistance of the segment of wire that represents the loss by  $y$, see Fig.\ref{sketch}. 

\begin{figure}[!b]

\begin{center}

\begin{pspicture}(-7,-5)(7,5)


\psset{arrowsize=0.2 2}

\pszigzag[coilarm=1.0, coilheight=0.15, linewidth=0.35mm]{-}(-4, 2.5)(0, 2.5)

\pszigzag[coilarm=1.0, coilheight=0.15, linewidth=0.35mm]{-}(0, 2.5)(4, 2.5)

\rput(0, 1.5){$(a)$}

\rput(-2, 3.35){$x$}

\rput(2, 3.35){$y$}

\rput(-4.5, 2.85){$V=V_0$}

\rput(4.5, 2.85){$V=0$}

\psdot(-4,2.5)

\rput(-5, 1.65){\small Newbiggin-by-the-Sea}

\rput(5, 1.65){\small Bottom of the ocean}
\psdot(4,2.5)

\pszigzag[coilarm=1.0, coilheight=0.15, linewidth=0.35mm]{-}(-4, -2)(0, -2)

\pszigzag[coilarm=1.0, coilheight=0.15, linewidth=0.35mm]{-}(0, -1)(4, -1)

\pszigzag[coilarm=1.0, coilheight=0.15, linewidth=0.35mm]{-}(0, -3)(4, -3)

\psline[linewidth=0.35mm](0,-1)(0,-3)

\psline[linewidth=0.35mm](4,-1)(4,-3)

\psline[linewidth=0.35mm](4,-2)(5,-2)

\psdot(5,-2)

\psdot(-4,-2)

\rput(0, -4){$(b)$}

\rput(-2,-1.15){$x$}

\rput(2,-0.25){$a-x$}

\rput(2,-2.25){$y$}

\rput(5.5, -1.65){$V=0$}

\rput(-4.5, -1.65){$V=V_0$}

\rput(-5, -2.65){\small Newbiggin-by-the-Sea}

\rput(6, -2.65){\small Sondervig}

\end{pspicture}

\caption{Circuit diagrams showing Heaviside's application of Blavier's method.}
\label{circuit2}

\end{center}

\end{figure}
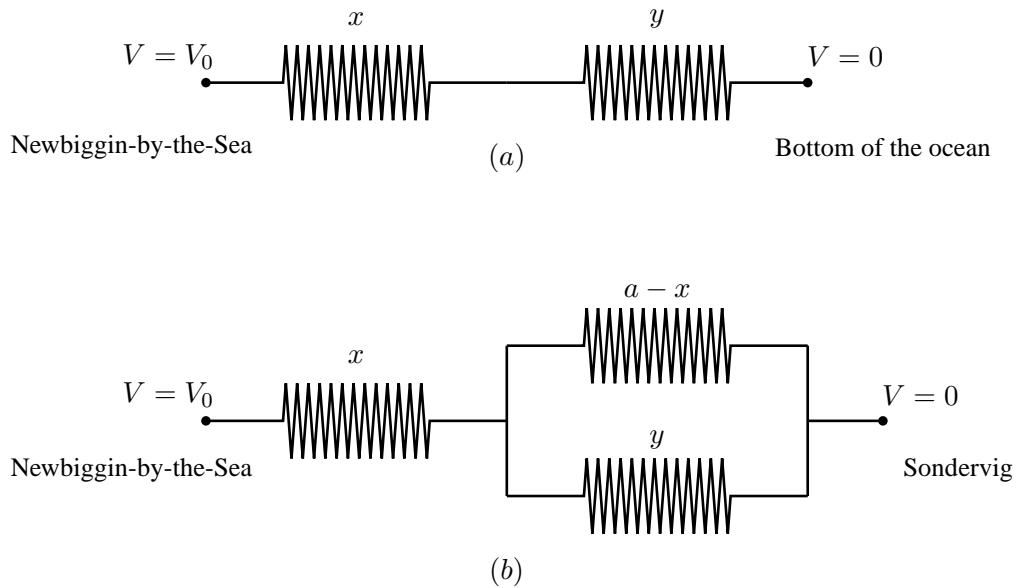

\noindent
Following Blavier's method, Heaviside notices that in order to measure the combination $b=x+y$, it suffices to interrupt the circuit at the Danish extremity and apply a known voltage and measure the current that flows through $x$ and $y$, see Fig.\ref{circuit2}  (a). In order to measure  $x$, it was necessary first to ground the Danish extremity and obtain $c$, the effective resistance as a function of $x$, $y$ and $a$. With the Danish extremity and the segment that represents the current loss grounded, the segments  $a-x$ and  $y$ are connected in parallel, see Fig.\ref{circuit2}  (b). 
The equivalent resistance of this combination is 
\[
R_{eq}  = \frac{{y\left( {a - x} \right)}}{{y + a - x}}.
\]
The effective resistance $c$ is then given by
\[
c = x + \frac{{y\left( {a - x} \right)}}{{y + a - x}},
\]
It follows that
\[
x^2  - cx + c\left( {a + y} \right) - ab = 0.
\]
Setting $y=b-x$, we obtain

\[
x^2  - 2cx + c\left( {a + b} \right) - ab = 0,
\]
the solution of which is

\[\label{x}
x = c \pm \sqrt {c^2  - c\left( {a + b} \right) + ab}  = c \pm \sqrt {\left( {c - a} \right)\left( {c - b} \right)} .
\]

\noindent
Heaviside data were $r=6\,\Omega$ per knot (in the context, $1$ knot =  $1$ nautical mile $=1852\,$m), $\ell=360\,$knots, $c=970\,\Omega $ and $b=1040\,\Omega$. It follows that  $a=2160\,\Omega$. Taking these values into equation (\ref{x}), we see that the physically acceptable solution is $x\approx 114\,$ knots or, approximately, $211\,$km from the English end. The history of the undersea telegraphic cable is a fascinating subject  and the interested reader may  get a good overview of it in, for example,  the book by Bodanis \cite{Bodanis}.

%
%
\section{Final remarks}

The problem of the subdivision of light is quite enlightening  in what concerns the utilisation of theoretical and applied physics in order to achieve technological and economical objectives. The second example, the problem of the location of the point of current leakage is a sample of the ingenuity of one of the most important physicist of the end of the 19th century. Though  dated, the two examples are simple and clear applications of textbook physics to concrete situations and can be discussed in a profitable way  in the classroom or assigned as independent study to the more interested student. 

Science in general, and physics in particular, are concrete manifestations of human societies which for several reasons reach an advanced level of cultural, social and economical development.  It is pedagogically healthy to remind ourselves and our students of this fact. This is best accomplished when we have suitable examples to help us in this task.

\section*{Acknowledgments}
The author wishes to thank his colleagues M V Cougo-Pinto and  V Soares for helping him improve the original manuscript. 



\begin{thebibliography}{99}
%
\bibitem{PN} Nahin P J 1988 Oliver Heaviside (Baltimore: John Hopkins)
%
%
\bibitem{Bourne} Bourn B 1996 The beginnings of electric street lighting in the City of London, Engineering Science and Education Journal, April 81-88
%
\bibitem{Darrigol} Darrigol O 2005 The Genesis of the Theory of Relativity,  S\'eminaire Poincar\'e \textbf{1} 1-22; see also 2000 Electrodynamics from Amp\`ere to Einstein, (Oxford: Oxford University Press)
%
\bibitem{Lorrain} Lorrain P,  Corson D R and Lorrain 2000 Fundamentals of Electromagnetic Phenomena (New York: Freeman)
%
\bibitem{Tipler} Tipler P A 1999  Physics for scientists and engineers 4th edition (New York: Freeman)
%
%
\bibitem{Bodanis} Bodanis D. 2005 Electric Universe: How Electricity Switched on the Modern World, (Three Rivers: Three Rivers Press )
\end{thebibliography}
\end{document}